# Growth of superconducting single-crystalline (Lu, Ca)Ba$_2$Cu$_3$O$_{7-\delta}$ whiskers

K. Deguchi [a, b, *], S. Ogawara [a], T. Okutsu [a, b], M. Nagao [a, d], T. Watanabe [a], Y. Mizuguchi [a, e], Y. Kubo [a], F. Tomioka [a, e], S. Ishii [a], S. Tsuda [a, c, e], T. Yamaguchi [a, e], M. Nagasawa [b], Y. Takano [a, b, e]

[a] National Institute for Materials Science, 1-2-1 Sengen, Tsukuba 305-0047, Japan

[b] Tokyo Denki University, 2-2 Kandanishiki-cho, Chiyoda-ku, Tokyo 101-8457, Japan

[c] WPI – MANA – NIMS, 1-1 Namiki, Tsukuba 305-0044, Japan

[d] University of Yamanashi, 4-4-37 Takeda, Kofu 400-8510, Japan

[e] JST, TRIP, 1-2-1 Sengen, Tsukuba 305-0047, Japan

**Abstract**

Single-crystalline (Lu, Ca)Ba$_2$Cu$_3$O$_{7-\delta}$ (Lu(Ca)123) whiskers have been successfully grown using the Te-doping method. X-ray diffraction patterns of Lu(Ca)123 whiskers showed sharp (0 0 $l$) peaks corresponding to REBa$_2$Cu$_3$O$_{7-\delta}$ phase (RE = rare earth elements). Transport measurements showed that the superconducting transition occurred at 83 K in the obtained whiskers.

## 1. Introduction

Since the discovery of a cuprate superconductor with a high transition temperature ($T_c$), many researchers have made a great effort to search for cuprate superconductors with favorable properties. Particularly, REBa$_2$Cu$_3$O$_{7-\delta}$ (RE123; RE = rare earth elements) has been studied actively, because of high potential for applications due to a high $T_c$ and a high critical current density under a high magnetic field. The series of RE123 whiskers have been grown using the Te- or Sb-doping method for RE = Y, La, Nd, Sm, Eu, Gd, Dy, Ho, Er, Tm, and Yb [1, 2]. However, there is no report about growth of superconducting single-crystalline LuBa$_2$Cu$_3$O$_{7-\delta}$ (Lu123) whiskers. Because ionic radius of Lu$^{3+}$ ion is a little smaller than the critical radius required to stabilize the RE123 structure, there has been a difficulty in synthesis of Lu123 phase [3, 4]. R. Pinto *et al.* synthesized single phase of the poly-crystalline Lu123 with Ca doping (Lu(Ca)123) [4]. In this paper, we report the growth of single-crystalline Lu(Ca)123 whiskers using the Te-doping method and their superconducting properties.

## 2. Experimental

To make precursor pellets, high purity powders of Lu$_2$O$_3$, CaCO$_3$, BaCO$_3$, CuO, and TeO$_2$ were mixed with the nominal compositions of LuCa$_x$Ba$_3$Cu$_3$O$_y$Te$_{0.5}$ ($x$ = 1.00, 1.25, 1.50, and 1.75). The mixed powders were calcined at 760, 790, and 820 ˚C for 10 h in air with three intermediate grindings. After calcinations, powders were pressed into pellets and set in an alumina boat. The precursor pellets were heated at 980 ˚C for 50 h in air, then cooled down to 880 ˚C at a rate of 1 ˚C/h, and finally cooled from 880 ˚C to room temperature for 20 h. During this heat treatment, whiskers were grown from the precursor pellets. We have characterized the obtained whiskers by X-ray diffraction with Cu-$K_\alpha$ radiation (XRD). Temperature dependence of resistivity was measured from room temperature down to 2 K by a four-probe method; silver paste electrodes were attached to the sample placed on an MgO substrate.

## 3. Results and discussions

We obtained whiskers from all pellets composed of LuCa$_x$Ba$_3$Cu$_3$O$_y$Te$_{0.5}$ ($x$ = 1.00, 1.25, 1.50, and 1.75). Figure 1 shows a photograph of a whisker grown from a pellet of $x$ = 1.00. Whiskers grown from the side of the pellet are shaped like needles. In contrast, plate-like whiskers were grown from the upper surface of the pellet. Figure 2 shows an XRD pattern of the whisker grown from a pellet of $x$ = 1.00. Five peaks were observed at $2\theta$ = 7.58°, 22.74°, 38.38°, 46.46°, and 54.80°, which were assigned to (0 0 1), (0 0 3), (0 0 5), (0 0 6), and (0 0 7), respectively. The $c$-axis length was obtained to be 11.720(1) Å from the peak positions. The $c$-axis lengths for $x$ = 1.25, 1.50, and 1.75 are 11.7191(5) 11.7192(5), and 11.719(1), respectively, showing negligible dependence on $x$. Figure 3 shows the temperature dependence of electrical resistivity for a whisker grown from the pellet of $x$ = 1.25. The superconducting transition of the whisker was observed at 83 K. The other whiskers grown from the pellets of $x$ = 1.00, 1.50, and 1.75 also showed superconducting transition at ~ 82K.

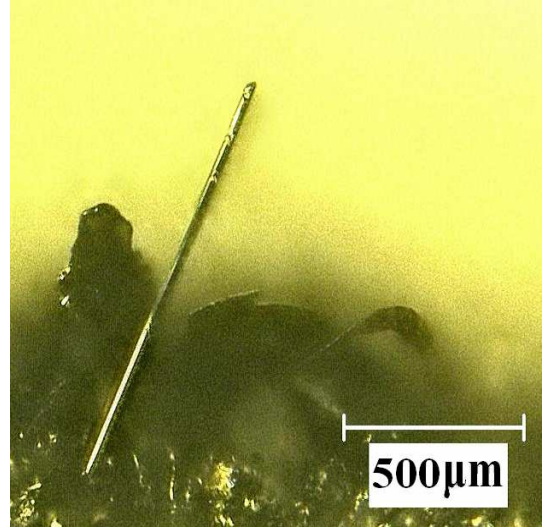

Fig.1. Photograph of a Lu(Ca)123 whisker grown from a precursor pellet composed of LuCa$_x$Ba$_3$Cu$_3$O$_y$Te$_{0.5}$ ($x$ = 1.00).

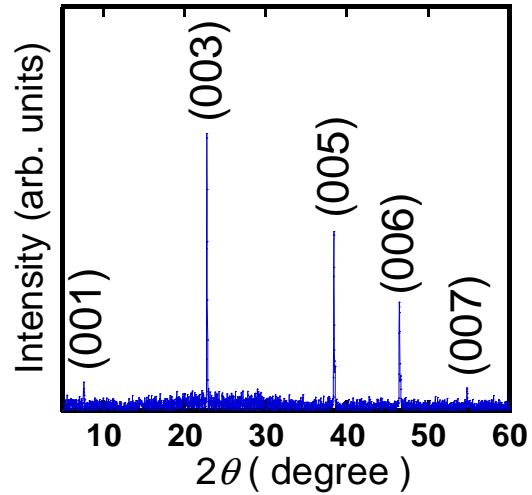

Fig.2. XRD pattern of a Lu(Ca)123 whisker grown from a precursor pellet composed of LuCa$_x$Ba$_3$Cu$_3$O$_y$Te$_{0.5}$ ($x$ = 1.00).

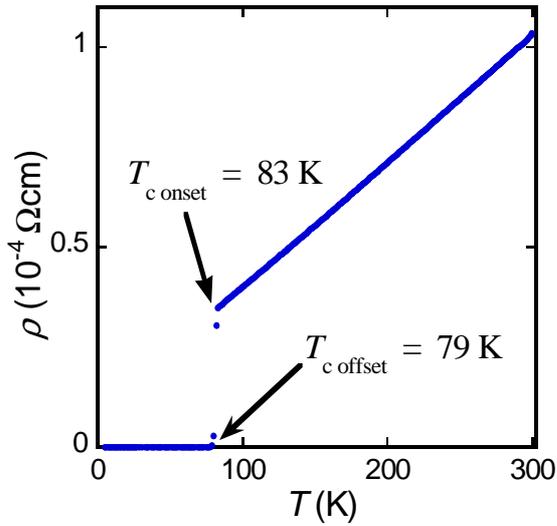

Fig.3. Temperature ($T$) dependence of resistivity ($\rho$) of a Lu(Ca)123 whisker grown from a precursor pellet composed of LuCa$_x$Ba$_3$Cu$_3$O$_y$Te$_{0.5}$ ($x$ = 1.25).

Thus, superconducting Lu(Ca)123 whiskers were successfully grown using the Te-doping method. We suppose that substitution of Ca for Lu increases the average ionic radii of the Lu sites, and stabilizes the Lu123 structure. The $c$-axis length and the transition temperature exhibited nearly no dependence on $x$ for $x$ = 1.00 ~ 1.75. These results suggest that the Ca concentrations of the whiskers are the same. The solubility range of Ca may be narrow in the Lu(Ca)123 whiskers grown by the Te-doping method.

## 4. Conclusion

We have successfully grown single-crystalline Lu(Ca)123 whiskers by the Te-doping method. A maximum $T_c$ of 83 K was observed in the whisker grown from a pellet composed of LuCa$_x$Ba$_3$Cu$_3$O$_y$Te$_{0.5}$ ($x$ = 1.25). However, the $c$-axis lengths and the transition temperatures for $x$ = 1.00, 1.50, and 1.75 differed only slightly from those for $x$ = 1.25. The solid solubility range of Ca may be narrow and the whiskers of $x$ = 1.00 ~ 1.75 may have the same concentration of Ca.

## Acknowledge

This work was partially supported by Grant-in-Aid for Scientific Research (KAKENHI).

## References

[1] M. Nagao, M. Sato, Y. Tachiki, K. Miyagawa, M. Tanaka, H. Maeda, K. S. Yun, Y. Takano, T. Hatano, Jpn. J. Appl. Phys. 43 (2004) L324.
[2] T. Okutsu, S. Ueda, S. Ishii, M. Nagasawa, Y. Takano, Physica C 468 (2008) 1929.
[3] E. Hodorowicz, S. A. Hodorowicz, H. A. Elick, J. Alloys Compounds 181 (1992) 445.
[4] R. Pinto, L. C. Gupta, Rajni Sharma, A. Sequiera, K. I. Gnanasekar, Physica C 289, (1997) 280.